\documentclass[10pt, twocolumn]{LML}
\usepackage{soul}

\tikzset{
photon/.style={decorate, decoration={snake,amplitude=2pt, segment length=5pt}, draw=black},
particle/.style={draw=black, postaction={decorate}, decoration={markings,mark=at position .5 with {\arrow[draw=black]{>}}}},
antiparticle/.style={draw=black, postaction={decorate}, decoration={markings,mark=at position .5 with {\arrow[draw=black]{>}}}},
gluon/.style={decorate, draw=black, decoration={coil,amplitude=4pt, segment length=5pt}},
goldstone/.style={draw=green,postaction={decorate},decoration={markings,mark=at position .5 with {\arrow[draw=blue]{>}}}},
external/.style={}
}

\begin{document}

\title{Phenomenology of a Fake Inert Doublet Model}

\author[a,b,c]{Damiano Anselmi\thanks{damiano.anselmi@unipi.it}}

\author[c]{Kristjan Kannike\thanks{kristjan.kannike@cern.ch}}

\author[c]{Carlo Marzo\thanks{carlo.marzo@kbfi.ee}}

\author[c]{Luca Marzola\thanks{luca.marzola@cern.ch}}

\author[c]{Aurora Melis\thanks{aurora.melis@kbfi.ee}}

\author[c]{Kristjan M\"u\"ursepp\thanks{kristjan.muursepp@kbfi.ee}}

\author[c]{Marco Piva\thanks{marco.piva@kbfi.ee}}

\author[c]{Martti Raidal\thanks{martti.raidal@cern.ch}}

\affil[a]{Dipartimento di Fisica ``E. Fermi'', Universit\`a di Pisa, Largo B. Pontecorvo 3, 56127 Pisa, Italy}
\affil[b]{INFN, Sezione di Pisa, Largo B. Pontecorvo 3, 56127 Pisa, Italy}
\affil[c]{Laboratory of High Energy and Computational Physics, NICPB, R\"avala 10, 10143 Tallinn, Estonia}
\date{\today}

\twocolumn[
\maketitle
\begin{onecolabstract}
We introduce a new way of modeling the physics beyond the Standard Model by considering fake, strictly off-shell degrees of freedom: the fakeons. To demonstrate the approach and exemplify its reach, we re-analyze the phenomenology of the Inert Doublet Model under the assumption that the second doublet is a fakeon. Remarkably, the fake doublet avoids the most stringent $Z$-pole constraints regardless of the chosen mass scale, thereby allowing for the presence of new effects well below the electroweak scale. Furthermore, the absence of on-shell propagation prevents fakeons from inducing missing energy signatures in collider experiments. The distinguishing features of the model appear at the loop level, where fakeons modify the Higgs boson $h\to\gamma\gamma$ decay width and the Higgs trilinear coupling. The running of Standard Model parameters proceeds as in the usual Inert Doublet Model case. Therefore, the fake doublet can also ensure the stability of the Standard Model vacuum. Our work shows that fakeons are a valid alternative to the usual tools of particle physics model building, with the potential to shape a new paradigm, where the significance of the existing experimental constraints towards new physics must necessarily be reconsidered.
\end{onecolabstract}
]

\saythanks

\section{Introduction} 
\label{sec:intro}

All experimental particle physics searches have not yet provided convincing evidence for the existence of physics beyond the Standard Model (SM). If taken at their face value, the current constraints indicate that new physics must appear far above the electroweak scale, or that it must be too weakly coupled to the SM to yield significant effects at the energy scales probed by experiments. In this work, we argue that this is not necessarily the case, because all experimental constraints, so far, have been interpreted under the assumption that new physics effects manifest in the form of conventional particles that, propagating on-shell, leave a detectable signature. 

We propose, instead, that new physics appears in the form of ``fake," strictly virtual degrees of freedom -- the fakeons~\cite{Anselmi:2017ygm,Anselmi:2018kgz}, which despite mediating new interactions very much the same way as usual particles do, are prevented from being on-shell. This precludes the possibility of observing the propagation of these ``particles" on any distance scale and, therefore, makes their direct observation unfeasible. This simple possibility forces a radical change of the paradigm used for the interpretation of experimental results in terms of new physics. In fact, many experimental constraints which invalidate most conventional models of new physics simply do not apply to the case of fakeons. Notably, the bounds from the precision $Z$-boson width measurements at LEP and all the direct LHC searches for new physics do not directly constrain the properties of fakeons. Therefore, as we show in the present work, new light electrically and weakly charged degrees of freedom may exist at low energy scales without contradicting any experimental result. We stress that although the direct observation of fake degrees of freedom is impossible, their presence can still be inferred from their virtual contributions in both tree-level and higher order processes. For the same reason, fakeons can be used to address the existing problems in particle physics through new virtual contributions.

The fakeons were originally proposed to tame the fatal ghost degrees of freedom that appear in quantum gravity~\cite{Anselmi:2017ygm} and in Lee-Wick theories~\cite{Anselmi:2017yux,Anselmi:2017lia,Anselmi:2018kgz} by making them strictly virtual through a different quantization prescription. On general grounds, the same treatment can be applied to any type of massive field, so that all massive particles that appear in beyond-the-SM (BSM) theories can be changed into fakeons through the same prescription.\footnote{Within the context of the SM, the current constraints exclude the possibility that all particles be fakeons with the exception of the Higgs boson~\cite{Anselmi:2018yct}.}

To-date, such a possibility is essentially unexplored, no fakeon model building has been developed and, consequently, the phenomenology of fakeons remains largely unknown. As a first step in this direction, the present work aims to begin the exploration of fakeon phenomenology in the context of BSM physics and demonstrate the reach of model building with fakeons. 

Concretely, we extend the SM by adding an extra scalar doublet $\Phi$, as is done in any two-Higgs-doublet model (2HDM)~\cite{Gunion:1989we,Branco:2011iw}, but assume that the new doublet is a fakeon. For simplicity, we restrict ourselves to the simplest 2HDM setup,  the Inert Doublet Model (IDM)~\cite{Deshpande:1977rw,Ma:2006km,Barbieri:2006dq,LopezHonorez:2006gr} (see Ref.~\cite{Belyaev:2016lok} for a recent review), where $\Phi$ does not acquire a vacuum expectation value (VEV) and does not couple to the SM quarks and leptons. To distinguish the framework from the usual IDM, we name it the fake Inert Doublet Model (fIDM). As mentioned before, although $\Phi$ is an $SU(2)_L$ doublet, the $Z$-pole constraints~\cite{Zyla:2020zbs} do not apply to $\Phi$, because $\Phi$ cannot be produced on-shell in the $Z$ decays. For the same reason, the conventional LHC searches that target BSM scalar bosons~\cite{Morvaj:2019ldy,Tao:2019qN} or Dark Matter through missing energy signatures~\cite{Chatrchyan:2014lfa,Aad:2015zva} also fail to constrain the fIDM. These properties allow  fakeon mass scales well below the electroweak scale, despite constraints from experimental searches especially sensitive to electrically charged degrees of freedom.

Still, the fake doublet can manifest itself in its virtual contributions to SM processes, targeted for instance by precision physics observables~\cite{Han:2020lta} or alter the renormalization evolution of parameters. In regard of this, we show that the outstanding problem concerning the SM vacuum instability~\cite{Degrassi:2012ry,Buttazzo:2013uya} due to the mass of the SM Higgs boson discovered at the LHC~\cite{Aad:2012tfa,Chatrchyan:2012ufa} being below the stability bound, is solved in the fIDM by the additional Higgs portal couplings to the fakeon doublet. In order to pinpoint possible signatures of the model, we also analyze the fake doublet contributions to the Higgs boson $h\to\gamma\gamma$ branching ratio, finding that it can be sizeable at low fakeon mass scales due to the absence of a tree-level contribution $h\to\Phi\Phi$ to the total Higgs boson decay width. Similarly, we compute the fakeon contribution to the yet unmeasured trilinear Higgs boson vertex. Importantly, the fIDM loop contributions to both of these processes inherently differ from those of the IDM by an imaginary part that, in the standard case, corresponds to the effect of on-shell particles in the loop. This contribution identically vanishes in the fakeon case.

The paper is organized as follows. In Sec.~\ref{sec:fakeons} we specify the Lagrangian of the model and review the most important phenomenological implications of fakeons. Sec.~\ref{sec:collider} is dedicated to the study of the model and offers detailed computations of its electroweak precision observables, as well as of the impact of the fake inert doublet on the Higgs boson phenomenology. We present our conclusions in Sec.~\ref{sec:summary} and in  Appendix~\ref{sec:fidm1loop} we list the new contributions to the counterterms that renormalize the trilinear Higgs boson vertex. 
Throughout the paper we present our results in terms of the relevant Passarino-Veltman functions $A_{0},B_{0}, B_{00}$ and $C_{0}$, where the subscripts refer to the irreducible and reducible cases.

\section{The fakeon prescription/projection} 
\label{sec:prescription}
Fake degrees of freedom rely on a non-standard quantization prescription and a projection that eliminates them from the Fock space of asymptotic states. The prescription ensures that the projection is consistent with the optical theorem at every loop order~\cite{Anselmi:2018kgz} and can be used alongside with the Feynman one  without spoiling unitarity, renormalizability and stability~\cite{Anselmi:2020tqo}. While in quantum gravity fakeons are necessary to have consistency~\cite{Anselmi:2017ygm,Anselmi:2018ibi}, in BSM physics they 
represent an interesting possibility, because their unusual features help evading some phenomenological constraints that apply to normal particles.
 
In more detail, the fakeon propagator formally reads
\begin{equation}\label{eq:fakeprop}
    i\frac{p^2-m^2}{(p^2-m^2)^2+\mathcal{E}^4}
\end{equation}
where $\mathcal{E}$ plays a role similar to that of the $\epsilon$ in the Feynman prescription. Inside loop diagrams, the propagator~\eqref{eq:fakeprop} must be used together with a set of prescriptions on the integration domains. A simpler, but equivalent formulation, at the level of the amplitude, is to treat the branch cuts associated with fakeons by means of a non analytic  Wick rotation, which is defined by the so called ``average continuation"~\cite{Anselmi:2017yux}
\begin{equation}\label{eq:AC}
    \mathcal{A}_{\text{f}}=\frac{1}{2}\left(\mathcal{A}_++\mathcal{A}_-\right).
\end{equation}
Here $\mathcal{A}_+$ and $\mathcal{A}_-$ are the amplitudes computed using the Feynman and anti-Feynman prescriptions, respectively. Notice that using (\ref{eq:fakeprop}) without the other instructions amounts to using the Cauchy principal value of $1/(p^2-m^2)$, which yields severe inconsistencies in loop diagrams~\cite{Anselmi:2020tqo}. In this paper we compute the relevant amplitudes by using formula~\eqref{eq:AC}. In all the situations we consider, the effect of the average continuation reduces to taking the real part of the total amplitudes. However, we stress that in general the fakeon prescription affects both the real and the imaginary parts of the amplitudes, since~\eqref{eq:AC} is to be applied to each fakeon threshold and not to the total amplitude.

As a result of the interplay between the prescription and the projection, fakeons always remain purely virtual and are truly removed from the Fock space of the theory. This is an intrinsic feature of fakeons, which radically differs from ordinary unstable particles, in this respect. For instance, although the $Z$ boson is virtual on the time scales probed in present experiments, we can in principle think of a setup where it is long lived enough to be detected as a final state (in the same way as muons are at the LHC). This is impossible for fakeons.

\section{A fake doublet extension of the SM} 
\label{sec:fakeons}

The fIDM is specified by the following Lagrangian

\begin{align}   \label{eq:lag0}
    \LG =& \LG_{\rm SM} + (D_\mu\Phi)^\dagger (D^\mu\Phi) -V \,, \\
    \label{eq:V}
    V=& - m_1^2 \modu{H}^2 + m_2^2 \modu{\Phi}^2  + \lambda_1 \modu{H}^4 + \lambda_2 \modu{\Phi}^4 
    \\
     +\lambda_3 & \modu{H}^2 \modu{\Phi}^2 + \lambda_4 \modu{H^\dagger \Phi}^2 +\frac{1}{2} \lambda_5 \left((H^\dagger \Phi)^2 + {\rm H.c.}\right),\notag
\end{align}
where $H$ is the SM Higgs doublet and
\begin{equation}\label{eq:phi}
    \Phi = 
    \begin{pmatrix}
    \phi^+ \\ \cfrac{\phi_{H} + i \phi_{A}}{\sqrt 2}
    \end{pmatrix}
\end{equation}
is a fakeon doublet that transforms in the $\{1, 2, 1/2\}$ representation of the $SU(3)_c\times SU(2)_L\times U(1)_Y$ gauge group. 

Eqs.~\eqref{eq:lag0} and \eqref{eq:V} are simply the Lagrangian of a 2HDM, where the Higgs doublet is even under a $\mathbb{Z}_2$ symmetry, while the fakeon doublet is odd under it, {\it i.e.}, $\Phi\rightarrow -\Phi$. Because of that, we have adopted the standard 2HDM conventions in denoting the scalar couplings and labelling the new degrees of freedom. For the sake of simplicity, we assume that the second doublet does not acquire a VEV and does not couple to the SM fermions. As a result, our Lagrangian is formally identical to that of the IDM~\cite{Deshpande:1977rw,Ma:2006km,Barbieri:2006dq,LopezHonorez:2006gr}. 
 Nevertheless, due to the different quantization prescription for the field $\Phi$, the fIDM gives rise to a phenomenology radically different from that of the usual IDM.

The fakeon field \eqref{eq:phi} is specified by the masses $m_i$ of its components,
their widths $\Gamma_i$ and the interactions encoded in Eq.~\eqref{eq:lag0} and \eqref{eq:V}. Accounting for the contributions of electroweak symmetry breaking, the fakeon masses are given by 
\begin{align}
\label{eq:m}
    m^2_i = & \, m_2^2 + \lambda_i v^2,\, i=\phi^\pm,\, \phi_H,\, \phi_A, \\
    \lambda_{\phi^\pm} = &  \, \lambda_3 ,\nn \\
    \lambda_{\phi_H} = & \, \lambda_3+ \lambda_4 +\lambda_5 ,\nn\\
    \lambda_{\phi_A} = &  \, \lambda_3+ \lambda_4 -\lambda_5, \nn
\end{align}
where $v$ is the VEV of the Higgs doublet. The widths $\Gamma_i$, as usual, are given by the imaginary parts of the corresponding fakeon self-energy diagrams. In the present case, the $\mathbb{Z}_2$ symmetry therefore forces these quantities to identically vanish. However, we remark that $\Gamma_i^{-1}$ in general cannot be interpreted as the fakeon lifetimes. 

Because of the novel quantization prescription, fakeons cannot directly contribute to the decay widths of the particles they couple to. Furthermore, the results of loop diagrams involving fakeons are generally different from the results of loop diagrams made of usual particles. The computation of physical processes may deviate from the usual one already at the tree level, because the associated cut diagrams may have fakeons in loops. Importantly, processes with the SM final states mediated by fakeons do not receive contributions from the (on-shell) pair production and subsequent decays of the fakeon mediators. A way to think about fakeon-mediated interactions is offered by the analogy with virtual pions that mediate the interactions in atomic nuclei.

Another interesting effect of fakeons is the violation of microcausality~\cite{Anselmi:2018tmf,Anselmi:2018bra}. In the static limit, the extension $r$ of the fakeon mediated virtual interaction is encoded in the Yukawa potential $1/r\, {\rm exp}(-m_\Phi r)$, which is exponentially suppressed by the fakeon mass. In the opposite limit, at high energies, $\sqrt{s} \gg m_\Phi,$ the propagation of virtual particles is restricted to small distances $1/\sqrt{s}$, similarly to, {\it e.g.}, the propagation of virtual photons $\gamma^*.$ Notice that the ``lifetime'' $\tau_\Phi=1/\Gamma_\Phi$ of fakeons does not play any role (assuming $\Gamma_\Phi \leq m_\Phi$), because fakeons cannot be produced on-shell. Therefore, the causality violation is always short-ranged~\cite{Anselmi:2019nie} and practically, inconsequential at the classical physics level.

To conclude the discussion of fakeon properties, we remark that in principle there exist two types of fakeons, distinguishable by the sign in front of their quadratic term~\cite{Anselmi:2018bra}. If the positive sign is chosen, the fakeon field has the quadratic term of a normal particle. When the opposite choice is made, a normal particle turns into a ghost, which carries negative energy and can be abundantly produced from the vacuum, leading to phenomenological disasters. In the case of fakeons, the negative quadratic term is harmless, because the fakeons cannot be produced on-shell. In the fIDM case the two options are physically equivalent, since there exists a map that identifies the two emerging theories through a simple redefinition of parameters. For this reason, we consider only the standard sign in Eq.~\eqref{eq:lag0}.

\section{Phenomenology of a fake inert doublet} 
\label{sec:collider}

The most important direct constraints on a generic 2HDM come from collider experiments~\cite{Branco:2011iw}. In particular, the $Z$-pole measurements exclude additional light particles of masses $m<m_Z/2$ that participate in electroweak interactions~\cite{Zyla:2020zbs}. However, because fakeons cannot be produced and propagate on-shell, our fake doublet components cannot contribute to the $Z$ width regardless of their mass scale and thus are perfectly allowed by this constraint. Similarly, the direct searches for BSM scalar bosons at the ATLAS~\cite{Morvaj:2019ldy} and the CMS~\cite{Tao:2019qN} experiments are not applicable to the fIDM. 

The second important aspect of fakeon phenomenology concerns the missing energy signatures in collider experiments, which are at the basis of Dark Matter searches~\cite{Chatrchyan:2011tn,Chatrchyan:2011zy,Aad:2012re,Chatrchyan:2014lfa,Aad:2015zva}. In the fIDM, the $\mathbb{Z}_2$ symmetry $\Phi\rightarrow -\Phi$ of Eq.~\eqref{eq:lag0} implies that every interaction vertex contains an even number of fakeon legs. Nevertheless, the fakeons cannot carry energy away, as they cannot be assigned to the final states, even in a first approximation. This statement is general and applies to any fakeon model. We also stress that in contrast to the standard IDM, the new fake degrees of freedom included in the fIDM do not provide a viable Dark Matter candidate, because they are necessarily virtual. 

Thirdly, as long as the $\mathbb{Z}_2$ symmetry is unbroken, the fake doublet cannot couple to the SM fermions. Therefore, within the fIDM the new physics effects occur only at the loop level. We now proceed to study the most relevant examples.

\subsection{Electroweak precision observables} 
\label{sec:EWPT}

The electroweak precision tests (EWPT) put strong constraints on the physics beyond the SM by targeting new contributions to the self-energies of electroweak gauge bosons. The radiative shifts induced by new physics can be written as 
\begin{equation}
\label{eq:vaccapol}
\delta\Sigma^{ab}_{\mu\nu}(q^2)=\Sigma_T^{ab}(q^2)\left(g_{\mu\nu}-\frac{q_\mu q_\nu}{  q^2}\right)+\Sigma_L^{ab}(q^2)\frac{q_\mu q_\nu}{  q^2}, 
\end{equation}
where $a,b=Z, W, \gamma$. We list in App.~\ref{sec:fidm1loop} the expressions obtained in the fIDM for $\Sigma_T^{ab}(q^2)$. In order to make contact with experiments, we consider the following precision observables
\begin{align}
    &\frac{m_W^2}{(m_W^2)_{\rm SM}} 
    = 
    1 - \frac{\alpha \Delta S+2c_W^2 \alpha \Delta T}{2 (c_W^2 -s_W^2)} 
 +\frac{\alpha \Delta U}{4 s_W^2},
    \\\nn
    \\
    &\frac{\Gamma^W_{\rm tot}}{(\Gamma^W_{\rm tot})_{\rm SM}}
    =
    1 - \frac{\alpha \Delta S+2c_W^2 \alpha \Delta T}{2 (c_W^2 -s_W^2)} +
 \frac{\alpha \Delta U}{4 s_W^2} + \alpha \Delta W,
    \\\nn
    \\
    &\frac{\Gamma(Z\to \nu\bar\nu)}{\Gamma(Z\to \nu\bar\nu)_{\rm SM}}
    =
    1 + \alpha \Delta T  + \alpha \Delta V,
    \\\nn
    \\
    &\frac{s^2_W(m^2_Z)}{(s^2_W)_{\rm SM}} 
    = 1+  \frac{\alpha \Delta S+4 s_W^2 c_W^2 \alpha \Delta T}{4 s_W^2 (c_W^2 -s_W^2)} 
 + \alpha\Delta X ,
\end{align}
written here in terms of the Peskin-Takeuchi parameters $S$, $T$ and $U$~\cite{Peskin:1990zt,Peskin:1991sw}, as well as the complementary $V$, $W$ and $X$ parameters~\cite{Maksymyk:1993zm}. The symbols $\alpha$, $c_W$ and $s_W$ denote the fine-structure constant and the cosine and sine of the Weinberg angle, respectively.
\begin{figure}[t]
\begin{center}
  \includegraphics[width=0.87\linewidth]{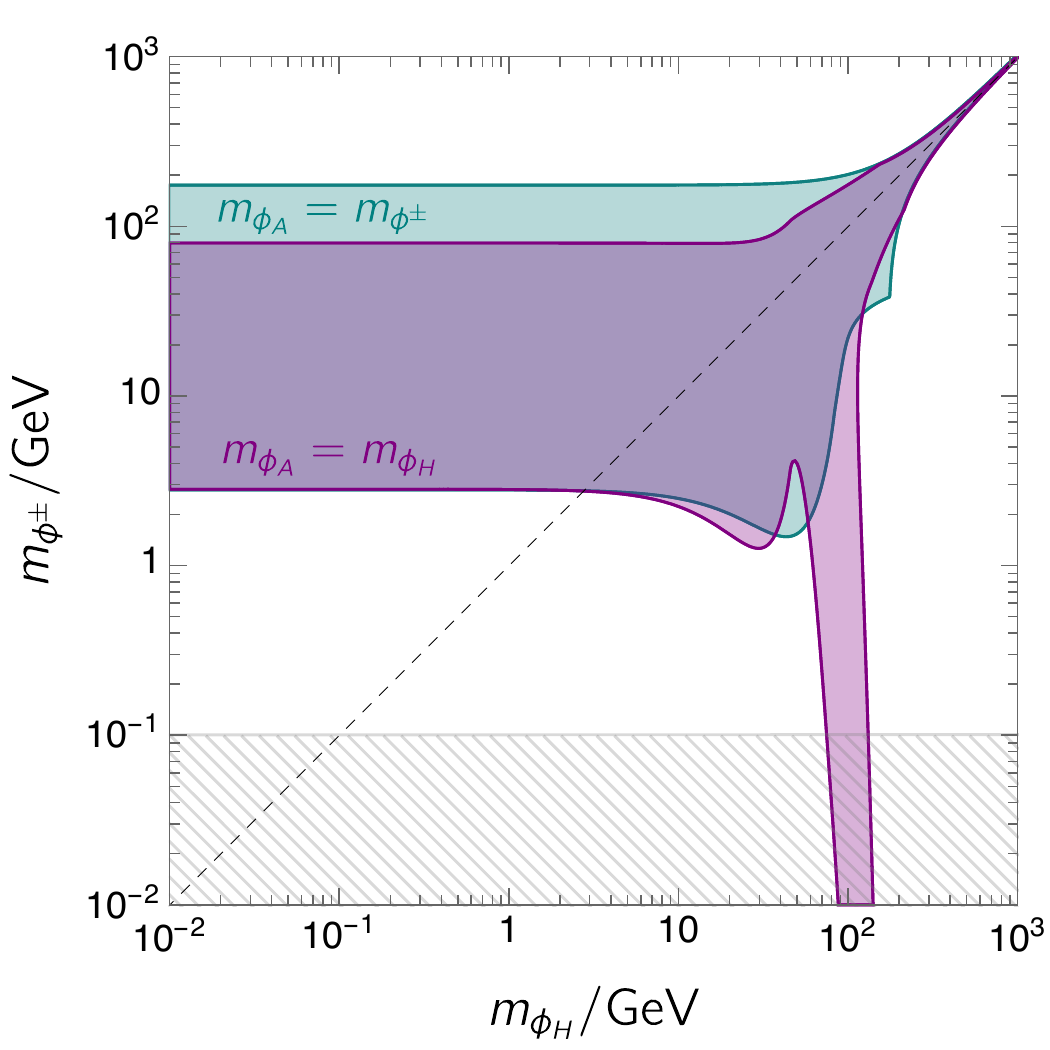}
\caption{\it The fake doublet masses allowed at the $3\sigma$ level from EWPT for the cases $m_{\phi_A} = m_{\phi^\pm}$ (green) and $m_{\phi_{A}} = m_{\phi_{H}}$ (purple). The hatched  region is excluded as explained in the text.}
\label{fig:ob1}
\end{center}
\end{figure}

Within the present framework, the oblique parameters receive new contributions sourced by the fake degrees of freedom appearing in the vacuum polarization diagrams of the electroweak gauge bosons. In terms of the expressions presented in App.~\ref{sec:fidm1loop}, we have: 
\begin{align}
    \frac{\alpha \Delta S}{4 s_W^2 c_W^2}
    &=
    \left[\frac{\Sigma^{ZZ}_T(m_Z^2)-\Sigma^{ZZ}_T(0)}{m_Z^2}\right]
    \\ \nn &
    -\frac{c_W^2-s_W^2}{s_W c_W}\Sigma^{'Z \gamma}_T(0)-\Sigma^{'\gamma \gamma}_T(0),
    \\\nn
    \\
    \alpha \Delta T 
    &= \frac{\Sigma^{WW}_T(0)}{m_W^2} - \frac{\Sigma^{ZZ}_T(0)}{m_Z^2},
    \\\nn
    \\
    \frac{\alpha \Delta U}{4 s_W^2}
    &=
    \left[\frac{\Sigma^{WW}_T(m_W^2)-\Sigma^{WW}_T(0)}{m_W^2}\right]
    \\ \nn& 
    -
    c_W^2\left[\frac{\Sigma^{ZZ}_T(m_Z^2)-\Sigma^{ZZ}_T(0)}{m_Z^2}\right]
    \\ \nn &
    -s_W^2\Sigma^{'\gamma \gamma}_T(0)-2 s_W c_W \Sigma^{'Z \gamma}_T(0),
    \\\nn
    \\
    \alpha \Delta V &= \Sigma^{'ZZ}_T(m_Z^2) - \left[\frac{\Sigma^{ZZ}_T(m_Z^2)-\Sigma^{ZZ}_T(0)}{m_Z^2}\right],
    \\\nn
    \\
    \alpha \Delta W &= \Sigma^{'WW}_T(m_W^2) - \left[\frac{\Sigma^{WW}_T(m_W^2)-\Sigma^{WW}_T(0)}{m_W^2}\right],
    \\\nn\\
    \alpha \Delta X &= - s_W c_W  \left[\frac{\Sigma^{Z \gamma}_T(m_Z^2)}{m_Z^2} -\Sigma^{'Z \gamma}_T(0)\right],
\end{align}
where a prime indicates differentiation with respect to the squared four-momentum $q^2$ of Eq.~\eqref{eq:vaccapol}. Notice that the above expressions of the Peskin-Takeuchi parameters differ from the ones usually considered in the Literature, as the latter are often simplified under the assumption that new physics be above the weak interaction mass scale. We also remark that in the case of fIDM, all oblique parameters are necessarily real regardless of the fakeon mass scale: vacuum polarization amplitudes containing circulating fakeons cannot give the imaginary contributions that, in the standard case, are associated to particle production via the cutting rules. 

The fake doublet masses allowed by the $3\sigma$ confidence intervals of the considered observables~\cite{Zyla:2020zbs} are shown in Fig.~\ref{fig:ob1}. The light purple region refers to a setup where the fake doublet neutral components are held degenerate in mass. In the area rendered in light green, instead, the mass of the pseudoscalar component is set to that of the charged one. The overlap of the two regions is rendered in a darker purple and the dashed line indicates a completely degenerate mass spectrum. All the considered mass splittings are obtained for perturbative values of the quartic couplings in Eq.~\eqref{eq:m}. As we can see, the oblique parameters do not forbid the presence of new fake degrees of freedom well below the electroweak scale. Possible mass hierarchies characterised by sub-GeV charged fakeons are partially excluded by the precise measurements of the fine-structure constant at the electron and muon mass scales. The corresponding bound is shown  by the hatched region in the plot.

\subsection{Stability of the electroweak vacuum} 
\label{sec:vac:stab}

In the SM, the self-coupling of the Higgs boson runs through zero to negative values at around $10^{10}$~GeV~\cite{Degrassi:2012ry,Buttazzo:2013uya}. This generates, via the Coleman-Weinberg mechanism, a global minimum at field values much above the electroweak scale, thereby destabilizing the SM vacuum. Models with new scalar degrees of freedom can address this potential vacuum instability problem thanks to the positive contributions to the Higgs boson quartic coupling $\beta$-function coming from the new portal couplings~\cite{Kadastik:2011aa,Khan:2015ipa}.

\begin{figure}[t]
\begin{center}
  \includegraphics{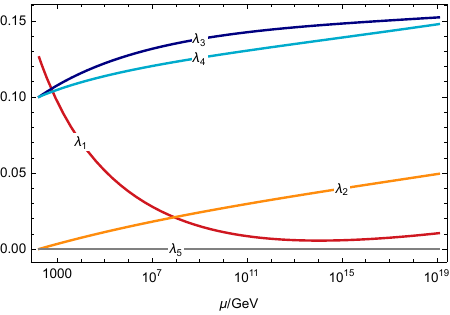}
\caption{\it The running scalar couplings as functions of the renormalization scale $\mu$ for the initial conditions $\lambda_{2} = \lambda_{5} = 0$, $\lambda_{3} = \lambda_{4} = 0.1$ given at the scale $m_{t}$.}
\label{fig:run}
\end{center}
\end{figure}

We use the {\tt PyR@TE 3} code~\cite{Sartore:2020gou} to derive the two-loop renormalization group equations of the couplings and take into account the dominant corrections from the top Yukawa coupling and the strong gauge coupling~\cite{Buttazzo:2013uya}. Setting all but one $\lambda_{i>1}$ to zero at the top quark mass scale, it is sufficient to have either $\lambda_{2} \gtrsim 0.26$ or $\lambda_{3} \gtrsim 0.19$ or $\lambda_{4} \gtrsim 0.16$ or $\lambda_{5} \gtrsim 0.28$ to keep the running Higgs self-coupling $\lambda_{1}$ positive and perturbative up to the Planck scale. With two non-zero new quartic couplings, values of order 0.1 are sufficient to ensure the stability of the electroweak vacuum. As an example, we show the running scalar couplings for $\lambda_{3} = \lambda_{4} = 0.1$ in Fig.~\ref{fig:run}. We stress that the RGEs for the fake doublet coincide with those of the usual IDM.

\subsection{Modified couplings of the Higgs boson} 
\label{sec:higgs}

Although the present experimental data have irrefutably shown the presence of a scalar CP-even particle of electroweak scale mass~\cite{Aad:2012tfa,Chatrchyan:2012ufa}, the presence of new physics in the dynamics of electroweak symmetry breaking and fermion mass generation has not been yet excluded. In regard of this, the couplings of the SM Higgs boson $h$ have been the main target of recent analyzes and future experimental probes~\cite{Abada:2019zxq,Benedikt:2018csr}, which aim at their complete and precise profiling.

While some of the properties of our model match those of the traditional IDM, the absence of fake scalars among the final states gives different predictions, which allow us to distinguish between the two cases. 
This is particularly evident for the $h\gamma \gamma$ effective coupling, which is responsible for the radiative decay $h \rightarrow \gamma \gamma$. This vertex is visible at the LHC as the final stage of $pp \rightarrow h \rightarrow \gamma \gamma$, and the experimental result is usually compared with the corresponding SM expectation through the diphoton rate
\begin{equation}
    R_{\gamma\gamma} = \frac{\sigma\left(pp \rightarrow h \rightarrow \gamma\gamma\right)_{\rm signal}}{\sigma\left(pp \rightarrow h \rightarrow \gamma\gamma\right)_{\rm SM}} .
\end{equation}
The latest ATLAS measurements report $R_{\gamma \gamma} = 1.03 \pm 0.11$~\cite{Aaboud:2018xdt,ATLAS:2019jst}, whereas the CMS finds $R_{\gamma \gamma} = 1.12 \pm 0.09$~\cite{Sirunyan:2018sgc,Sirunyan:2021ybb}.
In the present framework, the previous ratio translates into the measurement of the relative branching ratios
\begin{equation}
  R_{\gamma\gamma} \sim \frac{BR\left(h \rightarrow \gamma\gamma\right)_{\rm fIDM}}{BR\left(h \rightarrow \gamma\gamma\right)_{\rm SM}}, 
\end{equation}
which are sensitive both to a modification of the partial width $\Gamma_{h\rightarrow \gamma\gamma}$, as well as to a change in the total width of the Higgs boson that normalizes the ratios. In the usual IDM models, the latter includes new tree-level contributions from $h \rightarrow \phi_H \phi_H$, $h \rightarrow \phi_H \phi_A$ and $h \rightarrow \phi^+ \phi^-$, which, when kinematically allowed, suppress the resulting branching ratio. Within the fIDM, instead, the corresponding processes are always forbidden regardless of kinematics, because the scalars are fakeons.
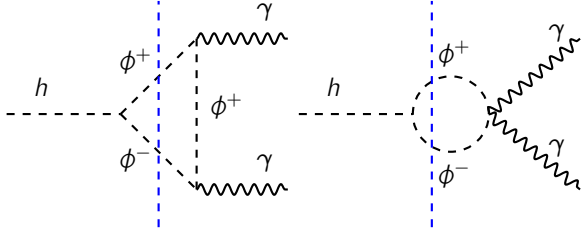
\begin{figure}[t]
\centering
\begin{tikzpicture}[thick,scale=1]
\draw[dashed] (-1.5, 0) -- node[black,above,xshift=-0.3 cm,yshift= 0.1cm] {$h$} (0,0);
\draw[dashed] (0,0) -- node[black,above,xshift=-0.3cm,yshift=-0.1cm] {$\phi^+$} (1,1);
\draw[blue, dashed] (0.5,-1.5) --  (0.5,1.5);
\draw[dashed] (0,0) -- node[black,above,xshift=-0.3cm,yshift=-0.4cm] {$\phi^-$} (1,-1);
\draw[dashed] (1,-1) -- node[black,above,xshift=0.4cm,yshift=-0.2cm] {$\phi^+$} (1,1);
\draw[photon] (1,1) -- node[black,above,xshift=0.3 cm,yshift=0.1cm] {$\gamma$} (2.2,1);
\draw[photon] (1,-1) -- node[black,above,xshift=0.3cm,yshift=0.1cm] {$\gamma$} (2.2,-1);
\end{tikzpicture}
\centering
\begin{tikzpicture}[thick,scale=1]
\draw[dashed] (-1.5, 0) -- node[black,above,xshift=-0.3 cm,yshift= 0.1cm] {$h$} (0,0);
\draw[dashed] (0.5,0) circle (0.5cm);  
\node at (0.55,0.8) [external]{$\phi^+$};
\node at (0.55,-0.8) [external]{$\phi^-$};
\draw[photon] (1,0) -- node[black,above,xshift=0.3 cm,yshift=0.33cm] {$\gamma$} (2.2,1);
\draw[photon] (1,0) -- node[black,above,xshift=0.3cm,yshift=-0.2cm] {$\gamma$} (2.2,-1);
\draw[blue, dashed] (0.25,-1.5) --  (0.25,1.5);
\end{tikzpicture}
\caption{\label{fig:h-2gamma}\it Fakeon-mediated contributions to $h\to \gamma\gamma$. The blue vertical lines denote the cuts that vanish for the fakeon prescription regardless of the kinematics. }
\end{figure}

A further difference between the two models arises in the computation of the relevant partial width, where the fakeon prescription prevents resonant loop contributions. In more detail, the general structure of the on-shell amplitude for $h\rightarrow \gamma \gamma$, as determined by Lorentz and gauge invariance, is given by 
\begin{equation}
\begin{split}
        M_{h\gamma\gamma} &= \frac{g^3  s_W ^2}{(4\pi)^2} F_{h \gamma \gamma}\times  \\ & \left(\epsilon\left(p_{1}\right)\cdot\epsilon\left(p_{2}\right) - \frac{2}{m_{h}^2} p_{1}\cdot \epsilon\left(p_{2}\right) p_{2}\cdot \epsilon\left(p_{1}\right) \right) ,
\end{split}
\end{equation}
where $\epsilon\left(p_{i}\right)$ are the polarization vectors of the final state photons with momenta $p_i$, $g$ is the weak coupling and $s_W$ is the sine of the Weinberg angle. The SM contribution to the form factor $F_{h \gamma \gamma}$ is dominated by the $W^{\pm}$ and top loops, given by
\begin{equation}
   F^{t}_{h\gamma\gamma} = \frac{8m_t^2}{3m_W}\left((m_h^2-4 m_t^2) C_0\left[0,0, m_h^2;m_t^2,m_t^2,m_t^2\right] - 2 \right), 
\end{equation}
\begin{align}
F^{W}_{h\gamma\gamma} = & m_W\Bigg(6 + \frac{m_h^2}{m_W^2} -6 (m_h^2-2 m_W^2)\nn
\\
&\times C_0\left[0,0,m_h^2;m_W^2,m_W^2,m_W^2\right]\Bigg) .
\end{align}
The new charged scalar states $\phi^{\pm}$ result in the additional term
\begin{equation} \label{hgagaFk}
\begin{split}
F^{\phi^{\pm}}_{h\gamma\gamma} = &  \frac{\lambda_3 m_W s_W^2}{\pi \alpha}\times \\ &\left(2 m_{\phi^\pm}^2\Re{C_0\left[0,0,m_h^2;m_{\phi^\pm}^2,m_{\phi^\pm}^2,m_{\phi^\pm}^2\right]} + 1 \right),
\end{split}    
\end{equation}
which differs from the IDM case by the absence of the imaginary part associated with the cuts of the three-point function shown in Fig.~\ref{fig:h-2gamma}. The discrepancy is due to the on-shell contribution of the particles in the loops, which identically vanishes in the fakeon case because of the quantization prescription of these fields.
Note that the Passarino-Veltman scalar two-point functions $B_0$ cancel out from the expression of the effective $h\gamma\gamma$ vertex.

\begin{figure}[t]
\begin{center}
\includegraphics[width=0.45\textwidth]{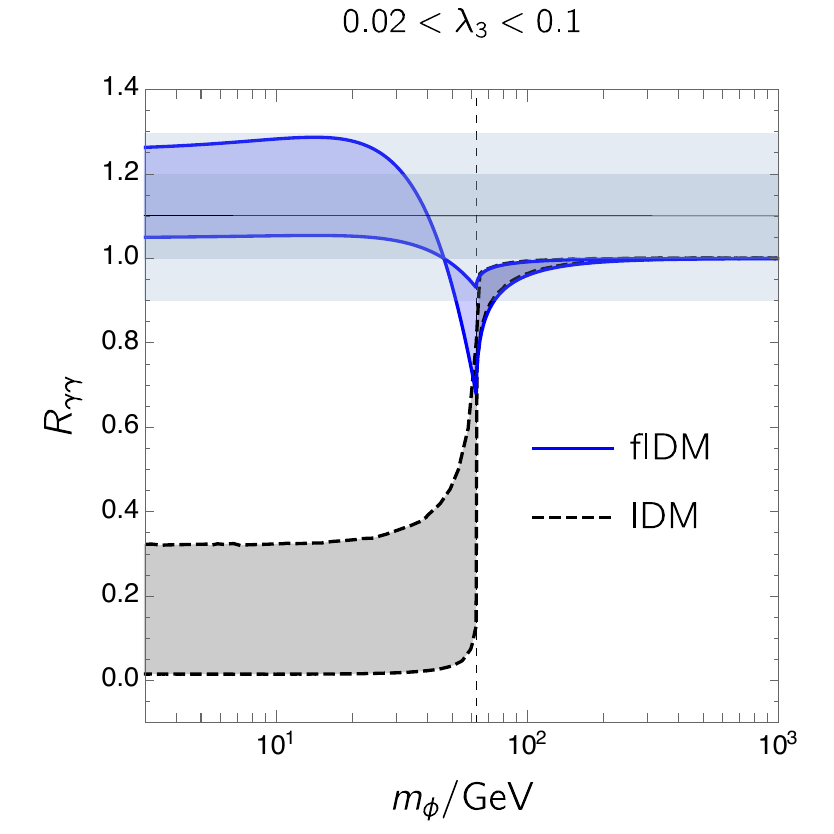}
\caption{\label{fig:hggresult}\it 
Predictions for $R_{\gamma\gamma}$ as functions of the degenerate doublet mass $m_\phi$ in the fIDM (solid lines) and the IDM (dashed lines). In both cases, the coupling $\lambda_3$ varies in the range $[0.02, 0.1]$. The region allowed by present experiments at the $2\sigma$  confidence level is shaded in light blue. Above the threshold, for $m_{\phi}>m_h/2$, the fIDM and the IDM predictions coincide. Below the threshold, for $m_{\phi}<m_h/2$, the fIDM values are larger because the total width of the Higgs boson does not receive a tree-level contribution from the production of fakeons.}
\end{center}
\end{figure}
\begin{figure}[t]
\begin{center}
\includegraphics[width=0.45\textwidth]{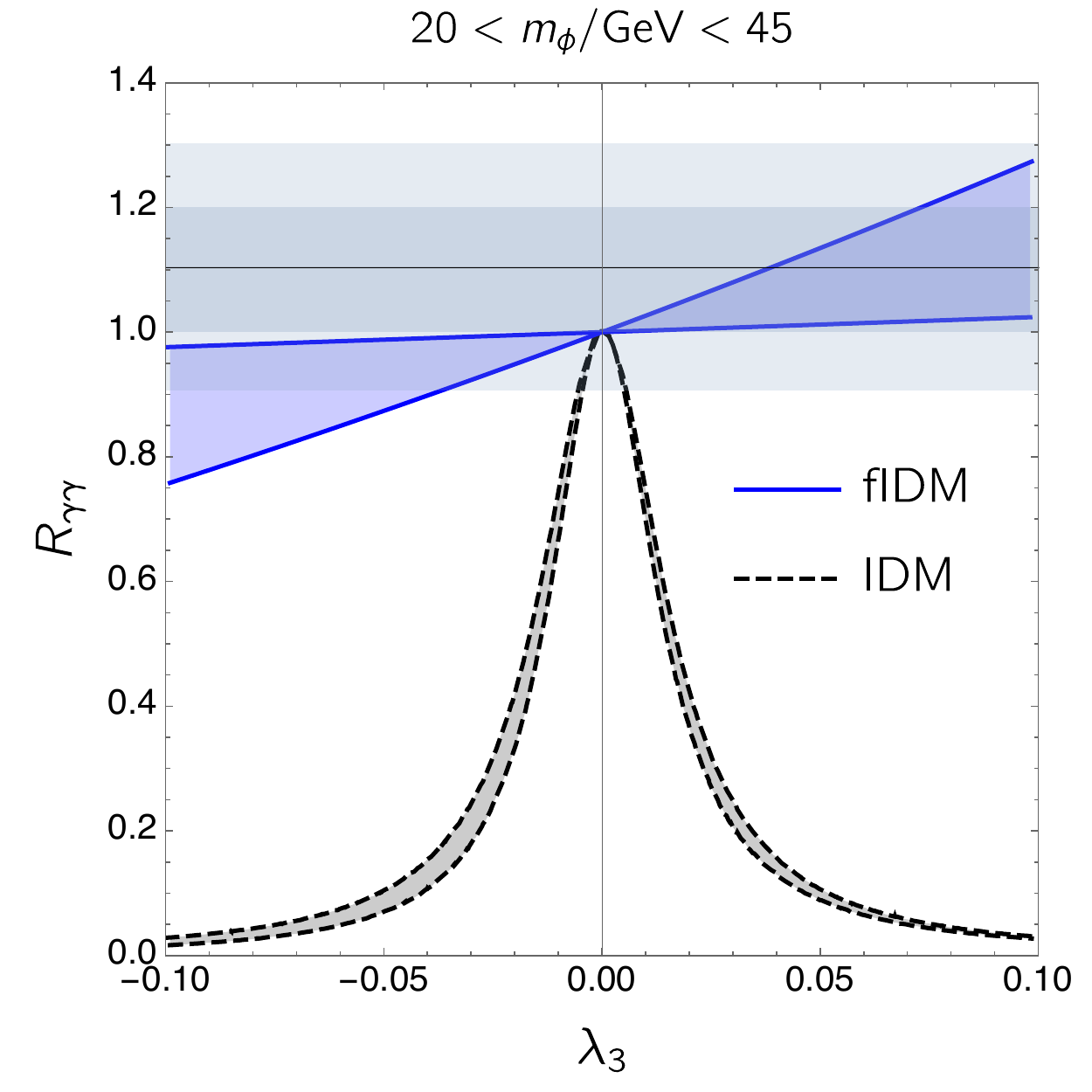}
\caption{\label{fig:hggresult2} \it 
Predictions for $R_{\gamma\gamma}$ as functions of the coupling $\lambda_{3}$ in the fIDM (solid lines) and in the IDM (dashed lines). The degenerate doublet mass varies below the threshold: $20<m_\phi/{\rm GeV} <45$. As in Fig.~\ref{fig:hggresult}, the fIDM contribution is larger, because the tree-level production of fakeons in the Higgs boson decays is forbidden.}
\end{center}
\end{figure}

In Fig.~\ref{fig:hggresult} we present numerical results for the predictions of $R_{\gamma\gamma}$ as functions of the degenerate doublet mass $m_{\phi}\equiv m_{\phi^\pm}=m_{\phi_H}=m_{\phi_A}$. The blue and gray regions are obtained by varying $\lambda_3$ in the range $[0.02, 0.1]$ within the fIDM and IDM, respectively. As we can see, in the fIDM case, the predicted values of $R_{\gamma\gamma}$ lie within the $2\sigma$ region selected by the current data on most of the considered parameter space. Importantly, the fIDM contribution (corresponding to the blue region) increases the diphoton rate for fakeon masses below $m_h/2$. The reason for this enhancement is that the total Higgs decay width is not suppressed by the tree-level production of new particles, which, instead, suppresses the corresponding IDM contribution (gray area). 
As for the absence of the imaginary part in Eq.~\eqref{hgagaFk}, we find that it does not affect the model prediction significantly. In fact, in either model,  new physics effects enter $\Gamma_{h\rightarrow \gamma\gamma}$ mainly through the interference with the dominant SM contribution, which is predominantly real.

That the fake doublet contribution to $h\to\gamma\gamma$ differs from that of an ordinary doublet with same mass and couplings is also demonstrated in Fig.~\ref{fig:hggresult2}, where we show $R_{\gamma\gamma}$ as a function of the coupling $\lambda_3$. The fakeon result is again measurable over the full range of parameters, whereas the IDM result is sizeable only for vanishing values of the $\lambda_{3}$ coupling. The two models are thus clearly distinguishable. In particular, within the usual IDM, obtaining $R_{\gamma\gamma}>1$ requires $\lambda_3<0$ and $m_h/2<m_\phi\lesssim 154\,{\rm GeV}$, and is never achievable for $m_\phi<m_h/2$. On the contrary, Figs.~\ref{fig:hggresult} and~\ref{fig:hggresult2} show that the enhancement of the diphoton rate is certainly possible in the case of the fIDM. We expect that the forthcoming high luminosity LHC measurements and the future $e^+e^-$~\cite{Abada:2019zxq} and/or $pp$~\cite{Benedikt:2018csr} colliders will reach enough precision to discriminate between the two models.

\begin{figure}[t]
\centering
\begin{tikzpicture}[thick,scale=1]
\draw[dashed] (-1.5, 0) -- node[black,above,xshift=-0.3 cm,yshift= 0.1cm] {$h^*$} (0,0);
\draw[dashed] (0,0) -- node[black,above,xshift=-0.3cm,yshift=-0.1cm] {$\Phi$} (1,1);
\draw[dashed] (0,0) -- node[black,above,xshift=-0.3cm,yshift=-0.4cm] {$\Phi$} (1,-1);
\draw[dashed] (1,-1) -- node[black,above,xshift=0.3cm,yshift=-0.2cm] {$\Phi$} (1,1);
\draw[dashed] (1,1) -- node[black,above,xshift=0.3 cm,yshift=0.1cm] {$h$} (2.2,1);
\draw[dashed] (1,-1) -- node[black,above,xshift=0.3cm,yshift=0.1cm] {$h$} (2.2,-1);
\draw[blue, dashed] (0.5,-1.5) --  (0.5,1.5);
\draw[red, dashed] (0,1) --  (1.5,0.5);
\draw[cyan, dashed] (0,-1) --  (1.5,-0.5);
\end{tikzpicture}
\centering
\begin{tikzpicture}[thick,scale=1]
\draw[dashed] (-1.5, 0) -- node[black,above,xshift=-0.3 cm,yshift= 0.1cm] {$h^*$} (0,0);
\draw[dashed] (0.5,0) circle (0.5cm);  
\node at (0.55,0.8) [external]{$\Phi$};
\node at (0.55,-0.8) [external]{$\Phi$};
\draw[dashed] (1,0) -- node[black,above,xshift=0.3 cm,yshift=0.3cm] {$h$} (2.2,1);
\draw[dashed] (1,0) -- node[black,above,xshift=0.3cm,yshift=-0.25cm] {$h$} (2.2,-1);
\draw[blue, dashed] (0.25,-1.5) --  (0.25,1.5);
\end{tikzpicture}
\caption{\label{fig:hhh}\it One-loop topologies contributing to the modification of the trilinear Higgs boson vertex within the fIDM. The dashed colored lines indicate the different thresholds that vanish identically for the fakeon quantization prescription.}
\end{figure}
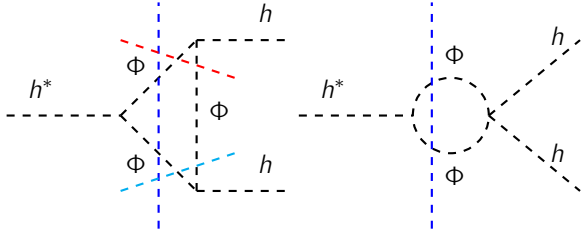

Another observable of interest for future collider experiments is the Higgs trilinear vertex $\lambda_{hhh}$, probed by 4$b$, 4$\tau$ or $2b2\tau$ final states. Clearly, the kinematic conditions over the three Higgs bosons involved in the vertex depend on the process under investigation. As a paradigmatic case, we compute the one loop diagrams shown in Fig.~\ref{fig:hhh} for an off-shell incoming Higgs boson and two outgoing on-shell particles, targeting the planned collider measurements. 
Because deviations as large as $100 \%$ of the SM prediction for the trilinear vertex are still experimentally allowed, we compute once again the new doublet contributions for both the fIDM and IDM. To this purpose, we define the quantity
\begin{equation}
   \rho_{hhh} = \frac{\lambda^{\rm (f)IDM}_{hhh}+\delta\lambda^{\rm (f)IDM}_{hhh}}{\lambda_{hhh}^{\rm SM}},
   \label{eq:lsmws}
\end{equation}
which measures the new physics loop contribution, in either the IDM or fIDM, normalized to the SM tree-level value $\lambda^{\rm SM}_{hhh}=-12 \lambda_1 m_W s_W/e$. As we show below, the prediction for $\rho_{hhh}$ is highly sensitive to potential imaginary parts sourced by the resonant loop contributions, corresponding to the thresholds indicated by the colored dashed lines in Fig.~\ref{fig:hhh}. Because the latter identically vanish within the fIDM, the observable can be used to distinguish between the two frameworks. We remark that the fakeon prescription can also affect the real part of the amplitude if kinematics allows all the internal lines to be simultaneously on-shell. This possibility is, however, precluded in our setup.

The computation of loop contributions to the trilinear Higgs boson vertex is particularly involved due to the workings of SM renormalization and constitutes, {\it per se}, a non-trivial check of the consistency of the theory. Indeed, the coupling of the $h$ trilinear vertex is not independent, so its renormalization is determined by the renormalization of the other parameters. To determine the one-loop unrenormalized amplitude of the process, we obtain the basic vertices of the theory by using {\tt FeynRules}~\cite{Alloul:2013bka} and generate the relevant diagram topologies with {\tt FeynArts}~\cite{Kublbeck:1990xc,Hahn:2000kx}. The resulting amplitude, as evaluated by {\tt FormCalc}~\cite{Hahn:1999mt}, then is 
\begin{align}
\label{eq:vhhh}
    & \lambda^{\rm (f)IDM}_{hhh}=  \nn \\ & \frac{m_W s_W}{8 \pi^2 e}\left( \frac{m_W^2 s_W^2}{\pi \alpha}\left(2 \lambda_3^3 C_0[m_h^2,m_h^2,q^2,m_{\phi^\pm}^2,m_{\phi^\pm}^2, m_{\phi^\pm}^2] \right.\right. \nn \\ & \left. + \lambda_{\phi_A}^3 C_0[m_h^2,m_h^2,q^2,m_{\phi_A}^2,m_{\phi_A}^2, m_{\phi_A}^2] \right. \nn \\ &  + \lambda_{\phi_H}^3 C_0[m_h^2,m_h^2,q^2,m_{\phi_H}^2,m_{\phi_H}^2, m_{\phi_H}^2] \Big)  \nn \\ & \left.
     +2 \lambda_3^3 \left(\frac{1}{2}B_0[q^2,m_{\phi^\pm}^2,m_{\phi^\pm}^2] +  B_0[m_h^2,m_{\phi^\pm}^2,m_{\phi^\pm}^2]\right) \right.
\nn \\ & \left.
     +\lambda_{\phi_A}^3 \left(\frac{1}{2}B_0[q^2,m_{\phi_A}^2,m_{\phi_A}^2] +  B_0[m_h^2,m_{\phi_A}^2,m_{\phi_A}^2]\right) \right.
\nn \\ & \left.
     +\lambda_{\phi_H}^3 \left(\frac{1}{2}B_0[q^2,m_{\phi_H}^2,m_{\phi_H}^2] +  B_0[m_h^2,m_{\phi_H}^2,m_{\phi_H}^2]\right)  \right),   
\end{align}
where $\lambda_{\phi_A}=\lambda_3+\lambda_4-\lambda_5$, $\lambda_{\phi_H}=\lambda_3+\lambda_4+\lambda_5$ .

In the on-shell renormalization scheme used in the SM, the Lagrangian parameters are related to the input parameters directly measured in precision experiments~\cite{Denner:1991kt}. The consistency of the theory requires that similar relations hold for the corresponding counterterms that subtract the UV-divergent parts of the loop corrections. In the case of the Higgs boson trilinear vertex, the dependence of the parameter on the Higgs boson quartic coupling, $\lambda_1 = \left(m_h^2 e^2\right) / \left(8 m_W^2 s_W^2\right)$, forces the corresponding counterterm to have the form~\cite{Denner:1991kt} 
\begin{align}
        & \delta\lambda^{\rm (f)IDM}_{hhh} = -\frac{ 3 e m_h^2}{2   s_W  m_W}\left( \delta Z_e - \frac{\delta   s_W  }{  s_W  } + \frac{\delta m_h^2}{m_h^2} \right. \nn \\ 
        & \left.  + \frac{e}{2   s_W  }\frac{\delta t}{m_W m_h^2} - \frac{1}{2}\frac{\delta m_W^2}{m_W^2} + \frac{3}{2}\delta Z_h \right),
        \label{eq:chhh}
\end{align}
determined by a set of radiative shifts received by the indicated SM couplings and masses. The above relation still holds true within the fIDM and IDM, provided that the additional radiative corrections sourced by the models are accounted for in the renormalization of the quantities appearing on the right-hand side. In Appendix~\ref{sec:fidm1loop}, we present the explicit expressions of these radiative corrections, which are shared by the two models, because they are unaffected by the fakeon prescription.

\begin{figure}[t]
\begin{center}
\includegraphics[width=0.48\textwidth]{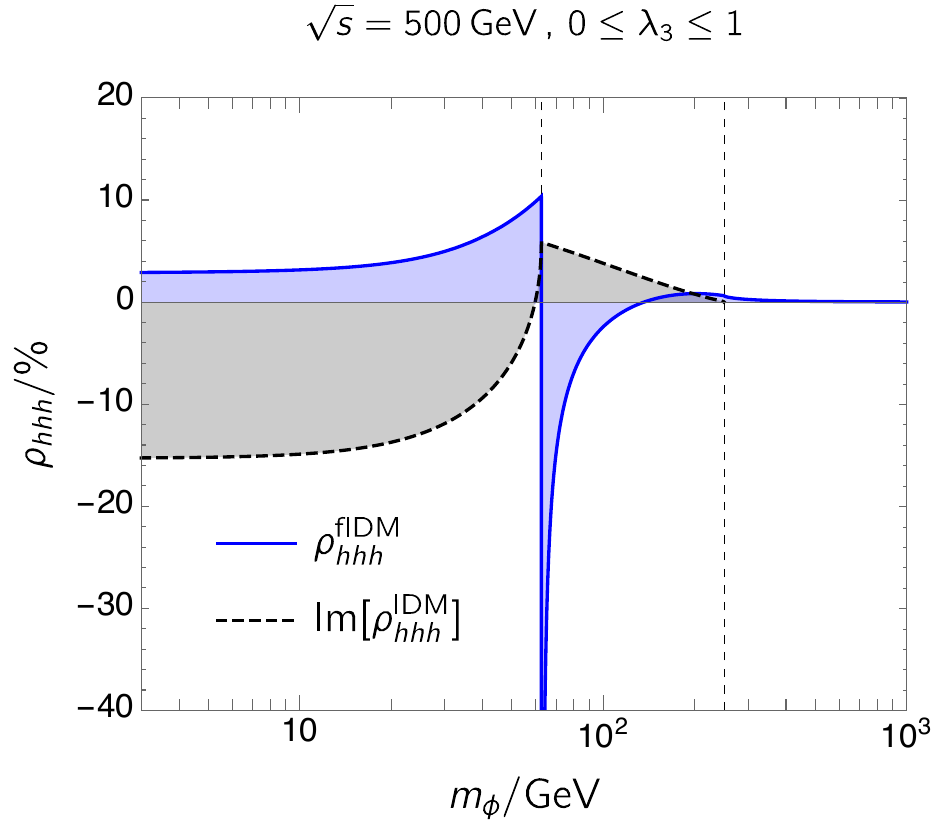}
\caption{\label{fig:hhhresult1} \em 
$\rho_{hhh}$ as a function of the fake doublet mass $m_{\phi}$ in the degenerate limit. The blue areas indicate the values of $\rho_{hhh}$ obtained for real contributions of the loop function in both the fIDM and IDM. Differently, the gray area shows the result of the imaginary part corresponding to the thresholds indicated in Fig.~\ref{fig:hhh}. This contribution vanishes within the fIDM but is generally present in the standard IDM. Notice the threshold $m_{\phi} = \sqrt{s}/2$, above which the two models give the same results.}
\end{center}
\end{figure}

The predictions obtained for the deviation of the trilinear Higgs boson vertex from the SM value are presented in Fig.~\ref{fig:hhhresult1}, which shows the dependence of $\rho_{hhh}$ on the new scalar masses in the degenerate limit. The plot was obtained for an example center-of-mass energy of $\sqrt{s}=500$ GeV by varying the quartic couplings in Eq.~(\ref{eq:vhhh}) within the indicated range and evaluating the involved loop function with {\tt LoopTools}~\cite{vanOldenborgh:1989wn}. The blue region indicates the results of the fIDM, which are given only by real contributions of the involved loop functions. The gray region, instead, represents the effect of the imaginary part of the amplitude that arises below the threshold $m_{\phi} = \sqrt{s}/2$ and generally enters the IDM prediction. Given the sizeable difference between real and imaginary contributions, the trilinear vertex can be used to discriminate between the fIDM and IDM. 

To conclude the discussion, we briefly comment on the consequences of possible violations of the $\mathbb{Z}_2$ symmetry imposed on the Lagrangian~\eqref{eq:lag0}. The simplest way to induce a $\mathbb{Z}_2$ violation in this model is to add Yukawa couplings between the fake doublet and the SM fermions, while keeping the $\mathbb{Z}_2$ charge assignments of all fermions positive. This would allow for fakeon-mediated processes resulting in the production of SM quarks and leptons. In the context of Higgs phenomenology, the $\mathbb{Z}_2$ violation would then yield multi-fermion final states via tree-level processes such as $H\to \phi^*\phi^*\to 4\ell,\, 4q,\, 2\ell 2 q,$ where $q$ and $\ell$ denote the SM quarks and leptons, respectively. The assessment of these peculiar signatures, which could help to constrain the properties of the $\mathbb{Z}_2$-violating fIDM, requires further analysis beyond the scope of this work.

\section{Summary} 
\label{sec:summary}

We have presented a new way of modelling physics beyond the SM based on strictly off-shell degrees of freedom: the fakeons. 

The phenomenological implications of this approach strongly diverge from those of more conventional model building. The different quantization prescription used for fakeons, in fact, forces the quanta of these fields to be purely virtual. Fakeons can thus mediate new interactions, just like ordinary particles, but cannot leave any direct signature in collider experiments. For this reason fakeons invalidate many constraints used to exclude ordinary models of new physics and clear the way for interesting and unexpected effects. 

As a concrete example, we have introduced the fake IDM extension of the SM -- the fIDM. Although the Lagrangian of the model is formally the same as that of the IDM, its phenomenology is radically different. A first outstanding distinction is that the new fake  doublet components can have masses well below the electroweak scale, in spite of the $Z$-pole precision measurements and other collider constraints. In fact, because the specific quantization prescription prevents fakeons from being on-shell, fakeons cannot be produced in $Z$-boson decays even if the new decay channel seems to be allowed by the kinematics. Furthermore, as in the case of the usual IDM, the $\mathbb{Z}_2$ symmetry at the basis of the model forbids Yukawa couplings between the new doublet and the SM fermions. Therefore, no processes with SM quark and lepton final states can be mediated at the tree level by the fake doublet. These properties ensure the absence of additional contributions to the $Z$-boson decay width and, similarly, allow fakeons to evade the typical collider constraints on new degrees of freedom. For instance, the virtual nature of fakeons prevents them also from carrying away energy at collider experiments and inducing missing energy signatures. 

Due to the $\mathbb{Z}_2$ symmetry and the impossibility of assigning fakeons to initial and final states, all new physics effects sourced by the fIDM occur necessarily at the loop level. To characterize the model, we have computed the electroweak precision constraints and found bounds on the mass splittings of the doublet components, while their absolute mass scale is unconstrained. We have also shown that the fake doublet interactions with the Higgs boson ensure the stability of the electroweak vacuum for perturbative values of the involved couplings. The distinguishing feature of the model is in its prediction for the $h\to\gamma\gamma$ branching ratio, which is generally enhanced owing to the lack of new tree-level contributions to the total decay width of the Higgs boson. Another observable of interest is the trilinear Higgs boson coupling, sensitive to the resonant loop contributions which are predicted to vanish identically within the fIDM. Both observables can be used to clearly distinguish the fIDM from the standard IDM. 

In conclusion, the introduction of fakeons into model building can yield a rich phenomenology at scales not necessarily larger than the electroweak one, without contradicting the available experimental results. The investigation of this paper is a first exploration of these scenarios and motivates further studies of this possibility for new physics.

\section*{Acknowledgements}
We thank Abdelhak Djouadi, Alessandro Strumia and Hardi Veerm\"ae for useful discussions. This work was supported by the Estonian Research Council grants PRG356, PRG434, PRG803, MOBTT86, MOBTT5 and by the EU through the European Regional Development Fund CoE program TK133 ``The Dark Side of the Universe". 

\begin{appendices}

\section{One-loop renormalization of the trilinear Higgs boson vertex in the fIDM}
\label{sec:fidm1loop}
 
In this Appendix we list the additional contributions to the self-energy diagrams $\Sigma$ and tadpole $T$ that determine the oblique parameters in Sec.~\ref{sec:EWPT} and the counterterm in Eq.~\eqref{eq:chhh} for the Higgs boson trilinear coupling. In the on-shell renormalization scheme, following Ref.~\cite{Denner:1991kt}, the explicit expressions of the counterterms $\delta Z_e$, $\delta m_W^2$, $\delta s_W$, $\delta m_h^2$, $\delta Z_h$ and $\delta t$ are given by
 \begin{align}
\delta Z_e & = \frac{1}{2}\frac{\partial \Sigma^{\gamma\gamma}_T(q^2)}{\partial q^2}\Bigg|_{q^2=0}-\frac{s_W}{c_W}\frac{\Sigma^{Z \gamma}_T(0)}{m^2_Z},\\[2pt]
\delta m^2_W & = {\rm Re}\left[\Sigma^{WW}_T(m^2_W)\right],\\[2pt]
\delta s_W & = \frac{c^2_W}{2 s_W} {\rm Re}\,\left[\frac{\Sigma^{ZZ}_T(m^2_Z)}{m^2_Z}-\frac{\Sigma^{WW}_T(m^2_W)}{m^2_W}\right],\\[2pt]
\delta m^2_h & = {\rm Re}\left[\Sigma^h (m^2_h)\right],\\[2pt]
\delta Z_h & =-{\rm Re}\left[\frac{\partial \Sigma^h(q^2)}{\partial q^2}\right]\Bigg|_{q^2=m^2_h},\\[2pt]
\delta t & = -T .
 \end{align}
The self-energies $\Sigma(q^2)$ due to the additional scalars are given by
 \begin{align}
\Sigma^{\gamma \gamma}_T(q^2) & = \frac{\alpha}{\pi}  \left(B_{00}[q^2, m_{\phi^\pm}^2, m_{\phi^\pm}^2] - \frac{1}{2}A_0[m_{\phi^\pm}^2] \right)    , 
 \end{align}
 \begin{align}
 \Sigma^{Z \gamma}_T(q^2) & = \frac{\left(1 - 2 s_W^2\right)\alpha}{2\pi c_W s_W}  \times  \nn \\
& \left(B_{00}[q^2, m_{\phi^\pm}^2, m_{\phi^\pm}^2] - \frac{1}{2}A_0[m_{\phi^\pm}^2] \right) ,
 \end{align}
 \begin{align}
&\Sigma^{Z Z}_T(q^2)  = \frac{\alpha}{4\pi c_W^2 s_W^2}\times \nn \\ 
& \left( B_{00}[q^2,m_{\phi_A}^2,m_{\phi_H}^2]-\frac{A_0[m_{\phi_H}^2]+A_0[m_{\phi_A}^2]}{4} \right)  \nn \\
& + \frac{\left(1 - 2 s_W^2\right)^2\alpha}{4\pi c_W^2 s_W^2}\left(B_{00}[q^2,m_{\phi^\pm}^2,m_{\phi^\pm}^2] - \frac{A_0[m_{\phi^\pm}^2]}{2} \right),
\end{align}
\begin{align}
\Sigma^{W W}_T(q^2)& = \frac{\alpha}{4\pi s_W^2}\times \nn \\
& \left( B_{00}[q^2,m_{\phi_H}^2,m_{\phi^\pm}^2]- \frac{A_0[m_{\phi_H}^2]+A_0[m_{\phi_A}^2]}{4} \right. + \nn \\
& \left.B_{00}[q^2,m_{\phi_A}^2,m_{\phi^\pm}^2] - \frac{A_0[m_{\phi^\pm}^2]}{2} \right),
\end{align}
\begin{align}
\Sigma^h(q^2) &= \frac{m_W^2 s_W^2}{32 \pi^2 } \times\nn\\& \left(2 \frac{\lambda_3^2}{\alpha \pi}  B_0[q^2,m_{\phi^\pm}^2,m_{\phi^\pm}^2] 
+ 2\lambda_3  A_0[m_{\phi^\pm}^2] \right.  \nn \\  
& \left. +\frac{\lambda_{\phi_H}^2}{\alpha \pi}  B_0[q^2,m_{\phi_H}^2,m_{\phi_H}^2] + \lambda_{\phi_H}  A_0[m_{\phi_H}^2]   \right.
 \nn \\  
& \left. +\frac{\lambda_{\phi_A}^2}{\alpha \pi}  B_0[q^2,m_{\phi_A}^2,m_{\phi_A}^2] + \lambda_{\phi_A}  A_0[m_{\phi_A}^2]   \right).
 \end{align}
 
New physics also shift the tree-level vacuum via the one-point diagrams $T$, entering the definition of $\delta t$ in the on-shell renormalization scheme, given by 
 \begin{align}
 T & = \frac{m_W s_W}{e 16 \pi^2}\times \nn \\
&\left( 2 \lambda_3 A_0[m_{\phi^\pm}^2] + \lambda_{\phi_H} A_0[m_{\phi_H}^2]+ \lambda_{\phi_A} A_0[m_{\phi_A}^2] \right).
 \end{align}

\end{appendices}


\bibliographystyle{JHEP}
\bibliography{fakeons}
 
\end{document}